\documentclass[]{jkpaper}

\usepackage{tikz}
\usetikzlibrary{decorations.markings}
\usetikzlibrary{decorations.pathreplacing}
\usetikzlibrary{arrows.meta}

\title{Unambiguous Phase Spaces for Subregions}
\author{Josh Kirklin}
\email{jjvk2@cam.ac.uk}
\institution{Department of Applied Mathematics and Theoretical Physics, Centre for Mathematical Sciences, University of Cambridge, Cambridge, UK}

\abstr{%
    The covariant phase space technique is a powerful formalism for understanding the Hamiltonian description of covariant field theories. However, applications of this technique to problems involving subregions, such as the exterior of a black hole, have heretofore been plagued by ambiguities arising at the boundary. We provide a resolution of these ambiguities by directly computing the symplectic structure from the path integral, showing that it may be written as a contour integral around a partial Cauchy surface. We comment on the implications for gauge symmetry and entanglement.
}

\bibliography{refs}

\begin{document}

\maketitleandtoc

\section{Introduction}
\label{Section: Introduction}
The Hamiltonian description of any classical physical theory consists of a phase space equipped with a symplectic structure, and a Hamiltonian function. The former is a specification of all the degrees of freedom in the theory, while the latter describes how these degrees of freedom evolve over time. Such a clear split between these two components is very useful when quantising the theory, as one may separately consider the quantum counterparts of each. The phase space is replaced by a Hilbert space, while the Hamiltonian function is replaced with a Hamiltonian operator.

This split between kinematics and dynamics is completely absent from the Lagrangian description of a classical theory. On the other hand, the Lagrangian approach has the advantage that one may often employ it in a manifestly covariant manner, which makes the symmetries of the theory easier to understand. However, the quantisation of a theory starting from its Lagrangian description is not a very well understood procedure, compared to starting from a Hamiltonian description. One must almost always first carry out some kind of Legendre transformation, in order to convert the Lagrangian description to a Hamiltonian one, and then proceed from there. One may argue that the Lagrangian path integral sidesteps this conversion. But such a path integral is usually only well-defined if we view it as an approximation to a Hamiltonian path integral.

It is still a relatively widely held misconception that in the course of such a conversion one must discard the covariance that makes the Lagrangian approach so attractive. This is most apparent in the canonical approach, the idea there being that one must take a snapshot of the physical system at some fixed time (which breaks covariance in the first instance), and then identify pairs of canonical conjugate variables in that snapshot, making a clear distinction between generalised coordinates and momenta (which breaks covariance in the second instance). One then computes the Hamiltonian in terms of these variables.

There is a different approach one can take. A point in the canonical phase space is the specification of a value for each coordinate and momentum at a fixed time. But the existence of the Hamiltonian implies that if one specifies values for the coordinates and momenta, one obtains a unique solution to the equations of motion. Similarly, given a solution to the equations of motion, one may deduce the values of the canonical variables at any moment in time. This means that there is a bijection between the canonical phase space and the space of solutions to the equations of motion. One may pullback the canonical symplectic structure to the space of solutions, which then makes the space of solutions a symplectic space isomorphic to the canonical phase space. This construction of the space of solutions and its symplectic structure is independent of the snapshot in time necessary for the canonical construction, and of the splitting between coordinates and momenta. Therefore, it is once more manifestly covariant. For this reason, the space of solutions is commonly known as the covariant phase space.

In the case of field theory, the covariant phase space formalism has its roots in~\cite{Peierls:10.2307/99080,DeWitt:1985bc,DeWitt:2003pm,Bergmann:PhysRev.89.4}, but was solidified in its modern form by~\cite{Crnkovic,300yearscrnkovicwitten,Zuckerman:1989cx}. It has since been explored in the work of~\cite{ashtekar1982,Lee:1990nz,Brown:1992br,Marolf:1993zk,Iyer:1994ys,Wald:1999wa,Barnich:2001jy,Hollands:2006zu} and many others. The formalism has found many applications, and recently it has been used to investigate symmetries of black hole spacetimes and aspects of the black hole information problem~\cite{Hawking:2016msc,Hawking:2016sgy,Haco:2018ske}. 

It is also relevant in holography, where the covariant phase space symplectic structure plays the role of the bulk dual to \changed[the boundary symplectic structure]{a natural symplectic structure on the space of boundary sources}~\cite{BELIN201971}. \changed{Further work in that context}~\cite{Belin:2018bpg} \changed{has investigated the relation between this symplectic structure and the volume of an extremal bulk slice, in particular revealing a connection to the complexity-volume conjecture}~\cite{Stanford:2014jda,Susskind:2014rva}.

Many of these studies rely on a common recipe for the symplectic structure. First, there is a procedure for deriving a certain differential form $\omega$ from the Lagrangian density. Then, one picks a Cauchy surface $\Sigma$. Finally, one integrates $\omega$ over $\Sigma$ to obtain the symplectic structure 
\begin{equation}
    \Omega=\int_\Sigma\omega.
    \label{Equation: intro symplectic 1}
\end{equation}
It is commonly assumed that one may also let $\Sigma$ be a partial Cauchy surface. Application of \eqref{Equation: intro symplectic 1} would then give a symplectic structure for the degrees of freedom in the subregion associated with that partial Cauchy surface, i.e.\ its domain of dependence. 

Unfortunately, this recipe suffers from a significant ambiguity. The form $\omega$ is only defined up to the addition of a certain class of exact forms. Under such a change $\omega \to \omega + \dd{\beta}$, the symplectic structure changes by a boundary integral, $\Omega \to \Omega + \int_{\partial\Sigma}\beta$. This will be described in more detail in Section \ref{Section: Review}. If $\Sigma$ has no boundary, then $\Omega$ is unmodified. But in many cases of physical significance $\Sigma$ does have a boundary (which may be either finite or asymptotic), and the ambiguity is a cause for genuine concern. Without a completely well-defined symplectic structure, the theory itself is ill-defined. 

Several approaches to dealing with this ambiguity have arisen. One might note that the ambiguity only affects physics at the boundary. Thus, if one is only concerned with physics deep in the interior of spacetime, one might argue that the ambiguity is irrelevant, so one may simply ignore it. However, in gauge theories this is untenable, due to the presence of non-local degrees of freedom which lead to correlations between the physics near the boundary and in the interior. Even if there is no gauge symmetry, this point of view is spoiled by the fact that very often we \emph{are} concerned with physics at the boundary. In fact in many cases the physics at the boundary is the main subject of interest. 

One example is the study of radiative degrees of freedom in asymptotically flat spacetimes. In this case $\Sigma$ asymptotes to spacelike (or null) infinity, and the radiative degrees of freedom contribute to the symplectic structure at this asymptotic boundary. In~\cite{Wald:1999wa} several natural conditions were imposed on the symplectic structure, based on physically sensible requirements for conserved charges. This somewhat reduces the boundary ambiguity, but does not altogether evade it.

Another example where the boundary physics is important arises in the case of a black hole spacetime, where it is natural to choose $\Sigma$ such that $\partial\Sigma$ intersects the event horizon -- with this choice, one is studying the physics on one side of the black hole. \changed[The above ambiguity leads to an uncertain value for]{One may show that certain quantities, such as} the black hole entropy\changed{, are unaffected by the ambiguity}~\cite{Jacobson:1993vj}, but there are \changed[also more complicated consequences]{non-trivial consequences of a more complicated nature}. For example, in~\cite{Haco:2018ske}, the charge algebra of large gauge symmetries\footnote{Large gauge transformations are gauge transformations whose action is non-trivial at the boundary of whichever region one is interested in.} at black hole horizons was studied. \changed{These charges are highly sensitive to the ambiguity.} The authors of that paper make a particular choice of boundary term, simply to make the large gauge transformations that they were interested in integrable (in the Hamiltonian sense). They (deliberately) provide no a priori justification for this prescription. 

Common to these approaches is the implicit belief that the recipe for the symplectic structure is completely correct, and that the boundary ambiguity must be fixed by additional, situation-dependent, considerations. 

In this paper, we will take an alternate viewpoint. We will argue that the boundary ambiguity is not actually present, and that it only arises because the recipe is incomplete. By a direct derivation from the Lagrangian path integral, we will show that the symplectic structure associated to $\Sigma$ is in fact given by a contour integral of $\omega$ around $\Sigma$. To be more precise, let $\mathcal{U}$ be any open submanifold of spacetime containing $\Sigma$. Then we find that the symplectic structure is given by 
\begin{equation}
    \Omega = \int_{\partial\mathcal{U}} \omega.
    \label{Equation: intro symplectic 2}
\end{equation}
Because $\partial(\partial\mathcal{U})=\emptyset$, the ambiguity $\omega \to \omega + \dd{\beta}$ is no longer an issue, as it does not result in a change in $\Omega$ as defined by \eqref{Equation: intro symplectic 2}. One may recover an expression resembling \eqref{Equation: intro symplectic 1} by taking the limit as $\mathcal{U}$ shrinks to contain only $\Sigma$, and by using certain causality conditions. The covariant phase space itself is also slightly modified in our approach. Whereas before it was given by the space of solutions to the equations of motion, we argue that it should instead be given by a space of field configurations which obey the equations of motion everywhere \emph{except} at $\Sigma$.

The outline of our derivation is as follows. First we define observables in the region associated with $\Sigma$ as observables which depend only on the field configuration on $\Sigma$. Then we compute the expectation value of the commutator of two such observables by inserting them into the path integral. Using the fundamental relation that arises in the classical limit between the commutator and the classical Poisson bracket, this allows us to obtain a Poisson structure for $\Sigma$. Finally, we invert this Poisson structure to obtain the symplectic structure.

\changed{Our result has implications whenever one needs to understand the physics that are relevant to an observer confined to a subregion, as for example the case in a black hole spacetime. We also expect that it is possible to apply our results in holography. In}~\cite{BELIN201971,Belin:2018bpg}\changed{, the authors considered the symplectic structure associated to the full boundary state. Our result provides a route for the extension of their analysis to the state in some subregion of the boundary. In particular, it could shed light on the complexity of the state in such a subregion.}

A short review of the pertinent features of the covariant phase space technique as it is usually employed may be found in Section \ref{Section: Review}; more extensive reviews appear in~\cite{Henneaux:1992ig,Compere:2018aar}. Our path integral derivation of the more accurate formulation described above is given in Section \ref{Section: Disambiguation}. We conclude with some remarks and speculation in Section \ref{Section: Conclusions} on the consequences of our results for gauge symmetry, edge modes, and entanglement.

\section{Covariant phase space review}
\label{Section: Review}

Let spacetime $\mathcal{M}$ be a $D$-dimensional Lorentzian manifold. A field configuration $\phi$ is a section of some bundle over $\mathcal{M}$. Let the space of all such field configurations be denoted $\mathscr{C}$. Let $\phi(s)$ be a smooth path in this configuration space. For each $s$ we have a field configuration $\phi(s)$, and varying $s$ leads to a smooth change in this field configuration. If one changes $s$ by an infinitesimal amount, the field configuration also changes infinitesimally, and this change is described by the tangent vector to the path $\phi(s)$. Thus, we can think of a linearised field variation to a particular field configuration as a vector in $\mathrm{T}\mathscr{C}$, the tangent bundle to $\mathscr{C}$. More generally, field-dependent linearised field variations may be viewed as sections of $\mathrm{T}\mathscr{C}$, i.e.\ vector fields on configuration space. We will use the notation $\delta\phi$ to refer to the vector field corresponding to the field variation $\phi \to \phi + \delta\phi$.

\subsection{Symplectic structure}

Field dynamics are described by an action $S[\phi]=\int_{\mathcal{M}}L[\phi] + S_{\partial\mathcal{M}}[\phi]$, where the Lagrangian density $L$ is a $D$-form that depends locally\footnote{In this paper, `local dependence' of some object $f$ on some other object $g$ means that for each $x\in\mathcal{M}$, $f(x)$ depends on $g$ only through expressions of the form $\mathrm{D} g(x)$, where $\mathrm{D}$ is some differential operator.} on $\phi$, and $S_{\partial\mathcal{M}}$ is a boundary term. Under an arbitrary field variation $\phi \to \phi+\delta\phi$, the change in the Lagrangian density (to linear order in $\delta\phi$) may be written 
\begin{equation}
    \delta L = \delta\phi \cdot E + \dd{\theta}.
    \label{Equation: Lagrangian variation}
\end{equation}
Here $E=\fdv{L}{\phi}$ is the Euler-Lagrange derivative of $L$, and the $\cdot$ denotes a summation over field indices. $\theta=\theta[\phi,\delta\phi]$ is a $(D-1)$-form that depends locally on $\phi$, and linearly locally on $\delta\phi$. Since $\theta$ depends linearly on the vector $\delta\phi$, it may be thought of as a 1-form on $\mathscr{C}$.

Using the above we may write the variation of the action as
\begin{equation}
    \delta S = \int_{\mathcal{M}} \delta\phi \cdot E + \int_{\partial\mathcal{M}}\theta + \delta S_{\partial\mathcal{M}}.
\end{equation}
The boundary term $S_{\partial\mathcal{M}}$ is chosen such that the latter two terms cancel each other, $\int_{\partial\mathcal{M}}\theta + \delta S_{\partial\mathcal{M}} = 0$. This often requires the use of supplementary boundary conditions at $\partial\mathcal{M}$. When this is done the variation of the action is just
\begin{equation}
    \delta S = \int_{\mathcal{M}} \delta\phi \cdot E.
\end{equation}
An on-shell field configuration is one for which $\delta S = 0$ for all $\delta\phi$. This is equivalent to the requirement that $E=0$, which are just the equations of motion.

The covariant phase space $\mathscr{P}$ is defined as the space of all on-shell field configurations. Any consistent phase space must be endowed with a symplectic structure, which we will do next.

Consider two field variations $\phi \to \phi + \delta_1\phi$ and $\phi \to \phi + \delta_2\phi$. Since these are vector fields on $\mathscr{C}$, we may define a third field variation given by their commutator $\delta_{12}\phi = [\delta_1\phi,\delta_2\phi]$. Now let
\begin{equation}
    \omega[\phi,\delta_1\phi,\delta_2\phi] = -\delta_1(\theta[\phi,\delta_2\phi]) + \delta_2(\theta[\phi,\delta_1\phi]) + \theta[\phi,\delta_{12}\phi].
    \label{Equation: symplectic definition}
\end{equation}
Let $\Sigma$ be a partial Cauchy surface in $\mathcal{M}$, and define
\begin{equation}
    \Omega[\phi,\delta_1\phi,\delta_2\phi] = \int_\Sigma \omega[\phi,\delta_1\phi,\delta_2\phi].
\end{equation}
$\Omega$ depends linearly on $\delta_1\phi,\delta_2\phi$, and is antisymmetric in these vectors. Therefore, it is a 2-form on $\mathscr{C}$. This 2-form may be pulled back to $\mathscr{P}$ -- from here on we will assume we have carried out this pullback. Then $\Omega$ is in fact a closed 2-form, as may be verified from the equation $\delta L = \dd{\theta}$ (which holds by definition on $\mathscr{P}$). We use $\Omega$ as the symplectic\footnote{Technically it is a \emph{pre}symplectic structure, because it may be degenerate. Any degenerate directions correspond to gauge symmetries or degrees of freedom causally disconnected from $\Sigma$, and may be eliminated by symplectic reduction.} structure on $\mathscr{P}$. 

So we end up with a phase space $\mathscr{P}$ equipped with a symplectic structure $\Omega$. This is the covariant phase space construction.

\subsection{Boundary ambiguities}

There are two boundary ambiguities in the above. The first is innocuous -- we may redefine $L \to L + \dd{K}$, $S_{\partial\mathcal{M}} \to S_{\partial\mathcal{M}} - \int_{\partial\mathcal{M}}K$ for some $(D-1)$-form $K$. This does not change the action, and hence leaves the dynamics invariant. But such a modification corresponds to a change $\theta \to \theta + \delta K$. Fortunately, the change in $\omega$ vanishes:
\begin{equation}
    \omega \to \omega - \underbrace{(\delta_1 \delta_2 K - \delta_2\delta_1 K - \delta_{12} K)}_{=0} = \omega
\end{equation}
The symplectic structure is unchanged, so we do not have to worry about this ambiguity.

The second ambiguity is far more serious. Equation \eqref{Equation: Lagrangian variation} only defines $\theta$ up to the addition of a closed form $k$ which is linearly locally dependent on $\delta\phi$. By the results of~\cite{Wald:2010:doi:10.1063/1.528839}, $k$ is exact, so let $k=\dd{\alpha}$. The corresponding change in $\omega$ is given by
\begin{equation}
    \omega \to \omega - \dd(\delta_1(\alpha[\phi,\delta_2\phi])-\delta_2(\alpha[\phi,\delta_1\phi])-\alpha[\phi,\delta_{12}\phi]),
    \label{Equation: omega ambiguity}
\end{equation}
and the symplectic structure changes by
\begin{equation}
    \Omega \to \Omega - \int_{\partial\Sigma} \delta_1(\alpha[\phi,\delta_2\phi])-\delta_2(\alpha[\phi,\delta_1\phi])-\alpha[\phi,\delta_{12}\phi].
\end{equation}
We see that if $\Sigma$ has a boundary (which is often the case in situations of physical relevance), the symplectic structure is not invariant under this ambiguity. 

This is something that we should be very concerned with. An ambiguous symplectic structure is a symptom of a sick theory. The cure is in the next section.

\section{Disambiguation of the covariant phase space}
\label{Section: Disambiguation}

\changed{For any given phase space, there are many possible symplectic structures one could choose. This choice is a purely classical one. However, not all symplectic structures will agree with the structure of the quantum theory implied by the path integral, and this gives us criteria which the `correct' symplectic structure ought to obey. For example, the symplectic structure gives us a Poisson bracket, which should agree with the quantum commutator in the semiclassical limit. Another possible criterion is agreement between the volume of the classical phase space implied by the Liouville measure associated with the symplectic form, and the dimension of the quantum Hilbert space.}

There is a method for recovering the correct symplectic structure from the semiclassical path integral. The outline is that one may compute the expectation value of the commutator of two observables by inserting the appropriate combination of corresponding operators into the path integral, and then taking the limit as the time-separation of these operators goes to zero. Using the relation between the quantum commutator and the classical Poisson bracket allows one to obtain a Poisson structure for the theory, which may then be inverted to obtain the symplectic structure. 

In this section we will employ this method, but restrict to observables which are accessible from within a subregion. We will assume that there is a sensible operator interpretation for the subregion observables. This will allow us to obtain a well-defined and unambiguous symplectic structure for that subregion. 

\changed{Our results will apply to a broad class of field theories with gauge symmetries. However, strictly speaking we should restrict to non-gravitational theories, i.e.\ those without diffeomorphism invariance. The restriction is necessary because a theory with diffeomorphism invariance does not have any local observables, and so the notion of the degrees of freedom in a subregion becomes much more subtle. In particular, if we are not careful, diffeomorphisms may move excitations in or out of the subregion under consideration. }

\changed{Nevertheless, we expect that it is possible to extend our analysis to theories of gravity, if one defines subregions in the correct way. In particular, subregions should be defined in a gauge-invariant manner (e.g.\ the exterior of the event horizon is certainly a gauge-invariant region of spacetime). Then we expect that similar results will apply. This would be interesting to verify, and is possibly connected to work in}~\cite{Donnelly:2016auv,Donnelly:2016rvo,Speranza:2017gxd,Camps:2018wjf}.

\subsection{Observables in a subregion}

A subregion is defined as a partial Cauchy surface $\Sigma\subset\mathcal{M}$. An observable $W[\phi]$ is a gauge-invariant function\footnote{It would be interesting to expand this analysis to include gauge-dependent observables by following ideas in~\cite{Marolf:1993af}. We leave this to future work.} on configuration space $\mathscr{C}$ that only depends on $\phi(x)$, and its derivatives normal to $\Sigma$, if $x \in \Sigma$. In the classical theory, for any given field configuration $\phi$, $W[\phi]$ is just a number which may be directly computed. The analogue of this number in the quantum theory is the expectation value $\expval*{\hat{W}}$ of some operator $\hat{W}$ associated with $W$. The classical limit is well-defined only if $\expval*{\hat{W}}$ converges to the classical number $W$ as $\hbar\to 0$.

We will now recall how this works for the semiclassical path integral, which for a quantum field theory with action $S$ is given by
\begin{equation}
    \mathcal{Z} = \int\Dd{\phi} \exp(iS/\hbar).
    \label{Equation: path integral}
\end{equation}
The integration is done over all field configurations which obey the boundary conditions. In the semiclassical limit $\hbar\to 0$, a saddlepoint approximation reveals that the path integral is dominated by configurations for which $S$ is extremised, i.e.\ those for which $\delta S = 0$ for an arbitrary choice of $\delta\phi$, or equivalently for which the equations of motion $E=0$ are obeyed. We assume that the boundary conditions are chosen such that there is only one solution to the equations of motion (up to gauge symmetry), which we will denote $\phi_0$. Then the path integral may be approximated by
\begin{equation}
    \mathcal{Z} = \exp(iS[\phi_0]/\hbar),
    \label{Equation: tree level}
\end{equation}
up to further factors which represent the contribution of quantum fluctuations away from $\phi_0$. Truncating these factors amounts to restricting to tree-level Feynman diagrams. \changed{This should be a good approximation whenever perturbation theory works, i.e.\ at weak coupling.} We will assume that \changed{for the theory we are considering} this truncation is a valid approximation, but it may be useful in the future to investigate the higher loop corrections in the following derivation.\footnote{\changed{Another possibility here would be to carry out a coherent state decomposition of the path integral. These are states which have a good classical limit, and so permit an approximation of the type in \eqref{Equation: tree level}. Corrections to \eqref{Equation: tree level} arise from the overlap between different coherent states, and this overlap also permits an interpretation in the classical limit. Such an approach would perhaps bear some similarity with the formalism in}~\cite{BELIN201971}.} \changed{It is not entirely clear whether our results extend to the case of strong coupling, but it may be possible to explore that regime by analytic continuation of the coupling constants.}

The expectation value $\expval*{\hat W}$ is defined by
\begin{equation}
    \expval*{\hat W} = \frac1{\mathcal{Z}}\int\Dd{\phi} W \exp(iS/\hbar).
\end{equation}
We may compute this expectation value by using a sourced path integral. First we introduce a sourced action $S(\epsilon) = S + \epsilon W$. The sourced path integral is then defined by
\begin{equation}
    \mathcal{Z}(\epsilon) = \int\Dd{\phi} \exp(iS(\epsilon)/\hbar),
\end{equation}
and the range of this integral is the same as for \eqref{Equation: path integral}. Then we clearly have
\begin{equation}
    \expval*{\hat W} = \left.-i\hbar\pdv{\epsilon} \log\mathcal{Z}(\epsilon)\right|_{\epsilon=0}.
    \label{Equation: sourced expval}
\end{equation}
In order to evaluate this expression, we need to know the value of $\mathcal{Z}(\epsilon)$ for small $\epsilon$, and this can be done by again using a saddlepoint approximation. The sourced path integral is dominated by configurations for which the sourced action $S(\epsilon)$ is extremised. For such configurations we have
\begin{equation}
    \delta (S+\epsilon W) = \int_{\mathcal{M}}\delta\phi \cdot E + \epsilon\delta W = 0
    \label{Equation: deformed eom}
\end{equation}
for all possible choices of $\delta\phi$. 

Suppose that $\phi_W$ is a field configuration obeying \eqref{Equation: deformed eom} and the boundary conditions. We assume that $\phi_W$ is unique, up to gauge symmetry. In the case that there is gauge symmetry, we will just pick one $\phi_W$ out of the possible gauge equivalent configurations -- all the following statements will be invariant with respect to this choice. We further assume that $\phi_W$ is a smooth function of $\epsilon$ satisfying $\left.\phi_W\right|_{\epsilon=0} = \phi_0$. Then $\phi_W$ defines a smooth path with parameter $\epsilon$ in the space of field configurations. Let $\delta_W\phi$ denote its tangent vector at $\phi_0$. 

Expanding in powers of $\epsilon$, the value of the sourced action at $\phi=\phi_W$ is given by
\begin{equation}
    S[\phi_W] + \epsilon W[\phi_W] = S[\phi_0] + \epsilon \int_{\mathcal{M}}\delta_W\phi \cdot \underbrace{E[\phi_0]}_{=0} + \epsilon W[\phi_0] + \order{\epsilon^2}.
\end{equation}
Using now the saddlepoint approximation 
\begin{equation}
    \mathcal{Z}(\epsilon) = \exp(iS[\phi_W]/\hbar)
\end{equation}
and \eqref{Equation: sourced expval}, we find
\begin{equation}
    \expval*{\hat W} = W[\phi_0].
\end{equation}
This expression is valid up to subleading in $\hbar$ corrections. Such corrections are negligible in the $\hbar\to 0$ limit. Therefore, given our assumptions above, the expectation value of $W$ attains its classical value in the classical limit, which is well-defined.

Before moving on to the next subsection, it will be useful to derive some further results concerning $\delta_W\phi$ and $S[\phi_W]$. Let $\mathcal{U}$ be a $D$-dimensional open submanifold of $\mathcal{M}$ such that $\Sigma\subset \mathcal{U}$, and consider first the case where $\delta\phi$ vanishes in $\mathcal{U}$. Then we have $\delta W = 0$, and may therefore write
\begin{equation}
    \qty[\int_{\mathcal{M}\setminus\mathcal{U}}\delta\phi \cdot E]_{\phi_W} = 0.
    \label{Equation: eom outside U}
\end{equation}
By the arbitrarity of $\delta\phi$, we may conclude that the equations of motion $E=0$ are obeyed outside of $\mathcal{U}$. Assuming this is true, \eqref{Equation: deformed eom} therefore reduces to
\begin{equation}
    \qty[\int_{\mathcal{U}}\delta\phi \cdot E + \epsilon \delta W]_{\phi_W} = 0.
    \label{Equation: deformed eom U}
\end{equation}
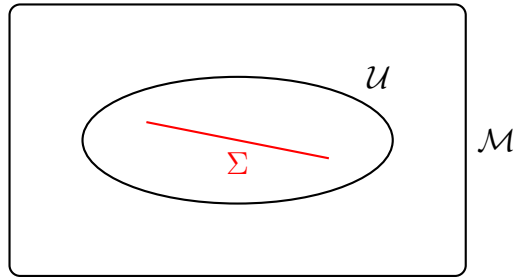
\begin{figure}[H]
    \centering
    \begin{tikzpicture}[thick,scale=1.2]
        \draw[rounded corners] (-0.5,0.5) rectangle (4.5,3.5);
        \draw (2,2) ellipse (1.7 and 0.7);
        \node at (3.55,2.7) {$\mathcal{U}$};
        \draw[red] (1,2.2) -- (3,1.8) node[midway, below] {$\Sigma$};
        \node[right] at (4.5,2) {$\mathcal{M}$};
    \end{tikzpicture}
    \caption{$\Sigma \subset \mathcal{U}\subset \mathcal{M}$. The action is deformed by a source term at the codimension one surface $\Sigma$.}
\end{figure}

One may write
\begin{equation}
    E[\phi_W] = E[\phi_0 + \epsilon\delta_W\phi] = \underbrace{E[\phi_0]}_{=0} + \epsilon \mathcal{S}(\delta_W\phi) + \order{\epsilon^2},
\end{equation}
where $\mathcal{S}$ is a linear differential operator characterising the linearised equations of motion. For example, in the case of a scalar field described by $E=\Box\phi = 0$, we have $\mathcal{S}=\Box$. Substituting this into \eqref{Equation: deformed eom U}, and considering only the $\order{\epsilon}$ term, one finds
\begin{equation}
    \delta W = -\int_{\mathcal{U}}\delta\phi \cdot \mathcal{S}(\delta_W\phi),
    \label{Equation: delta W}
\end{equation}
where this equation is understood to hold at $\phi=\phi_0$.

Next, we will make the connection with the formalism described in Section \ref{Section: Review}. One may substitute \eqref{Equation: Lagrangian variation} into \eqref{Equation: deformed eom U} to obtain
\begin{equation}
    \qty[\int_{\mathcal{U}} \delta L - \int_{\partial\mathcal{U}}\theta[\delta\phi] + \epsilon \delta W]_{\phi_W} = 0.
    \label{Equation: deformed eom symplectic potential}
\end{equation}
$\delta L$ and $\theta$ may be viewed as 1-forms in configuration space, since they depend linearly on $\delta\phi$. By considering configuration space Lie derivatives with respect to $\delta_W\phi$, we can obtain expansions about $\phi_0$ of these objects, in powers of $\epsilon$. One finds
\begin{equation}
    \qty[\delta L]_{\phi_W} = \qty[\delta L + \epsilon \delta_W\delta L + \order{\epsilon^2}]_{\phi_0},
\end{equation}
and
\begin{equation}
    \big[\theta[\delta\phi]\big]_{\phi_W} = \qty[\theta[\delta\phi] + \epsilon \qty(\delta_W(\theta[\delta\phi]) - \theta\big[\comm{\delta_W\phi}{\delta\phi}\big]) + \order{\epsilon^2}]_{\phi_0},
\end{equation}
where $\comm{\delta_W\phi}{\delta\phi}$ is the configuration space commutator of the two vectors $\delta_W\phi,\delta\phi$. Substituting these into \eqref{Equation: deformed eom symplectic potential}, and considering only the $\order{\epsilon}$ term, one obtains
\begin{equation}
    \delta W= \int_{\partial \mathcal{U}} \qty(\delta_W(\theta[\delta\phi]) - \theta\big[\comm{\delta_W\phi}{\delta\phi}\big]) - \int_{\mathcal{U}}\delta_W\delta L,
    \label{Equation: deformed eom symplectic potential 2}
\end{equation}
which holds at $\phi=\phi_0$. Noting that
\begin{align}
    \delta_W\delta L = \delta \delta_W L &= \delta\Big(\delta_W\phi \cdot E + \dd(\theta[\delta_W\phi])\Big) \\
                                         &= \delta_W\phi \cdot \mathcal{S}(\delta\phi) + \dd{\big(\delta(\theta[\delta_W\phi])\big)},
\end{align}
we may write \eqref{Equation: deformed eom symplectic potential 2} as
\begin{equation}
    \delta W= \int_{\partial \mathcal{U}} \qty(\delta_W(\theta[\delta\phi]) - \delta(\theta[\delta_W\phi]) - \theta\big[\comm{\delta_W\phi}{\delta\phi}\big]) - \int_{\mathcal{U}}\delta_W\phi\cdot \mathcal{S}(\delta\phi).
    \label{Equation: deformed eom symplectic potential 3}
\end{equation}
Referring back to \eqref{Equation: symplectic definition}, one recognises the first integrand as $-\omega[\delta_W\phi,\delta\phi]$. Therefore, using \eqref{Equation: delta W} to eliminate $\delta W$, we find
\begin{equation}
    \int_{\partial\mathcal{U}} \omega[\delta_W\phi,\delta\phi] = \int_{\mathcal{U}} \Big(\delta\phi \cdot \mathcal{S}(\delta_W\phi) - \delta_W\phi \cdot \mathcal{S}(\delta\phi)\Big).
    \label{Equation: omega S}
\end{equation}

Finally, we will obtain an expression for the $\order{\epsilon^2}$ term in the value of the sourced action at $\phi=\phi_W$. We have
\begin{align}
    S[\phi_W] + \epsilon W[\phi_W] &= S[\phi_0] + \frac12\epsilon^2\qty[\delta_W\qty(\int_{\mathcal{M}}\delta_W\phi\cdot E)]_{\phi_0} + \epsilon \Big(W[\phi_0] + \epsilon \delta_W W[\phi_0]\Big) + \order{\epsilon^3} \\
                                   &= S[\phi_0] + \epsilon W[\phi_0] +\epsilon^2\qty[\frac12\int_{\mathcal{M}}\delta_W\phi\cdot\mathcal{S}(\delta_W\phi) + \delta_W W]_{\phi_0} + \order{\epsilon^3} \\
                                   &= S[\phi_0] + \epsilon W[\phi_0] - \frac12\epsilon^2\qty[\int_{\mathcal{U}}\delta_W\phi\cdot\mathcal{S}(\delta_W\phi)]_{\phi_0} + \order{\epsilon^3}.
    \label{Equation: extremal sourced action}
\end{align}
In the last line, we used \eqref{Equation: eom outside U} to restrict the integral to $\mathcal{U}$, and then used \eqref{Equation: delta W} with $\delta\phi=\delta_W\phi$ to eliminate $\delta_W W$.

\subsection{Operator composition and the Poisson bracket}

We need a Poisson bracket for observables on $\Sigma$. Such a bracket should agree with the commutator in the classical limit. To be precise, for any two observables $A,B$ on $\Sigma$, we require
\begin{equation}
    \frac1{i\hbar}\expval*{\comm*{\hat A}{\hat B}\!} \to \pb{A}{B} \qq{as} \hbar\to0.
    \label{Equation: Poisson/commutator}
\end{equation}
This can be taken as the definition of the Poisson bracket.

In order for the commutator to make sense, we need a notion of operator ordering. In the path integral, this is implemented by `causal' boundary conditions, i.e.\ those such that $\delta_W\phi$ only has support in $J^+(\Sigma)$, for any observable $W$ on $\Sigma$. Here $J^+(\Sigma)$ is the causal future of $\Sigma$, i.e.\ the set of points in $\mathcal{M}$ which can be reached by following a future-directed\footnote{One must assume that $\mathcal{M}$ has a time-orientation.} causal curve starting in $\Sigma$. 

We will assume that our boundary conditions are causal. Operator ordering then translates directly to time ordering. An insertion of $\comm{A}{B}$ into the path integral really means an insertion of the combination $A(t)B(-t)-B(t)A(-t)$, where $A(t),B(t)$ are versions of $A,B$ which have been displaced a certain amount in time $t$. One takes the limit $t\to0$ from above, after having carried out the path integration\footnote{It is important that this limit takes place after the path integration. If one were to take the limit first, one would find a vanishing commutator, since $A,B$ are only c-numbers in the path integral, so
\begin{equation*}
    \lim_{t\to 0} \big(A(t)B(-t) - B(t)A(-t)\big) = AB-BA = 0.
\end{equation*}
When doing the path integral first, an $\order{1/t}$ number of paths will have significant contributions, so a non-zero quantity will result from the $t\to0$ limit.
}.

Clearly we will need a notion of time-displacement for the observables $A,B$ on $\Sigma$. To that end, let $\Sigma(t)\subset\mathcal{M}$ be a smooth 1-parameter family of partial Cauchy surfaces such that $\Sigma(0)=\Sigma$, and such that
\begin{equation}
    \Sigma(t_1)\subset J^+(\Sigma(t_2)) \qq{if} t_1>t_2.
\end{equation}
This condition says that $\Sigma(t_1)$ is to the future of $\Sigma(t_2)$ whenever $t_1>t_2$. Now let $A(t),B(t)$ be a pair of observables on each $\Sigma(t)$, smooth in the parameter $t$, and such that $A(0)=A$, $B(0)=B$. $A(t),B(t)$ are the time-displaced observables.

It is not clear that the way in which this time-displacement should be chosen to happen is unique, i.e.\ that there is a unique choice of the surfaces $\Sigma(t)$ and the time-displaced observables $A(t),B(t)$. The choice is clearly not completely free, as there are a number of consistency conditions that must be obeyed for the operator interpretation to make sense, one of which we will take advantage of below. Nevertheless, the final expression we will obtain appears to be independent of this choice.

By the above considerations, we may write the expectation value of $AB$ as
\begin{equation}
    \expval*{\hat A \hat B} = \lim_{t\to 0}\int\Dd{\phi} A(t)B(-t) \exp(iS/\hbar).
\end{equation}
As in the previous subsection, we can compute this expectation value using a sourced path integral. First let
\begin{equation}
    S(\sigma,\tau) = S + \sigma A(t) + \tau B(-t)
\end{equation}
be the sourced action, and define the sourced path integral as
\begin{equation}
    \mathcal{Z}(\sigma,\tau) = \int\Dd{\phi} \exp(iS(\sigma,\tau)/\hbar).
\end{equation}
Then we have
\begin{equation}
    \expval*{\hat A \hat B} = -\hbar^2\lim_{t\to 0} \qty[\frac1{\mathcal{Z}(\sigma,\tau)}\pdv[2]{\mathcal{Z}(\sigma,\tau)}{\sigma}{\tau}]_{\sigma=\tau=0}
    \label{Equation: expval AB}
\end{equation}
We will again use a saddlepoint approximation for this computation, writing
\begin{equation}
    \mathcal{Z}(\sigma,\tau) = \exp(iS_{\text{extremal}}(\sigma,\tau)/\hbar),
\end{equation}
where $S_{\text{extremal}}(\sigma,\tau)$ is the extremal value of the sourced action. The most efficient way to compute this quantity is to use the results of the previous subsection, with the substitutions 
\begin{align}
    \epsilon W &\to \sigma A(t) + \tau B(-t), \\
    \epsilon \delta_W\phi &\to \sigma \delta_{A(t)}\phi + \tau\delta_{B(-t)}\phi.
\end{align}
When one does this, it is important to ensure that $\mathcal{U}$ is chosen to contain both $\Sigma(t)$ and $\Sigma(-t)$. Using these substitutions in \eqref{Equation: extremal sourced action}, one finds
\begin{equation}
    S_{\text{extremal}}(\sigma,\tau) = S + \sigma A(t) + \tau B(-t) 
    -\frac12\int_{\mathcal{U}}\Big(\sigma \delta_{A(t)}\phi + \tau \delta_{B(-t)}\phi\Big)\cdot \mathcal{S}\Big(\sigma \delta_{A(t)}\phi + \tau \delta_{B(-t)}\phi\Big) + \dots
\end{equation}
where the right-hand side should be evaluated at $\phi=\phi_0$, and the ellipsis contains terms of cubic order and higher in $\sigma,\tau$. At this point, it is simple to apply \eqref{Equation: expval AB}, and one obtains
\begin{equation}
    \expval*{\hat A \hat B} = \lim_{t\to 0} \Big(A(t)B(-t)
    + \frac{i\hbar}{2}\int_{\mathcal{U}}\big[ 
        \delta_{A(t)}\phi \cdot \mathcal{S}(\delta_{B(-t)}\phi) +
    \delta_{B(-t)}\phi \cdot \mathcal{S}(\delta_{A(t)}\phi)\big]\Big) + \order{\hbar^2}.
\end{equation}
We have $\lim_{t\to 0} A(t)B(-t) = AB$. Also, by our assumptions about causality, an operator insertion at $t$ cannot affect an observation at $-t$, so we have
\begin{equation}
    \int_{\mathcal{U}}\delta_{A(t)}\phi\cdot\mathcal{S}(\delta_{B(-t)}\phi) = -\delta_{A(t)}B(-t) = 0.
\end{equation}
Thus, we may write
\begin{equation}
    \expval*{\hat A \hat B} = AB + \frac{i\hbar}{2}\lim_{t\to 0}\int_{\mathcal{U}}\delta_{B(-t)}\phi \cdot\mathcal{S}(\delta_{A(t)}\phi) + \order{\hbar^2}.
    \label{Equation: expval AB 2}
\end{equation}

A useful result arises from the following consistency condition:
\begin{equation}
    \expval*{\hat W}^* = \expval*{\hat{W}^\dagger},
\end{equation}
i.e.\ the complex conjugate of the expectation value is the expectation value of the Hermitian conjugate. We will assume that $A,B$ are both real observables, which means that their corresponding quantum operators are Hermitian. Therefore,
\begin{equation}
    \expval*{\hat A \hat B}^* = \expval*{\hat{B}^\dagger \hat{A}^\dagger} = \expval*{\hat B \hat A}.
    \label{Equation: AB consistency}
\end{equation}
Taking the complex conjugate of \eqref{Equation: expval AB 2}, we have
\begin{equation}
    \expval*{\hat A \hat B}^* = AB - \frac{i\hbar}{2}\lim_{t\to 0}\int_{\mathcal{U}}\delta_{B(-t)}\phi \cdot\mathcal{S}(\delta_{A(t)}\phi) + \order{\hbar^2}.
\end{equation}
Also, swapping $A$ and $B$ in \eqref{Equation: expval AB 2} yields
\begin{equation}
    \expval*{\hat B \hat A} = AB + \frac{i\hbar}{2}\lim_{t\to 0}\int_{\mathcal{U}}\delta_{A(-t)}\phi \cdot\mathcal{S}(\delta_{B(t)}\phi) + \order{\hbar^2}.
\end{equation}
By \eqref{Equation: AB consistency}, the right-hand sides of the above two equations are equal. As a consequence we find
\begin{equation}
    \lim_{t\to 0} \qty(\delta_{A(-t)}B(t) + \delta_{B(-t)}A(t)) = -\lim_{t\to 0}\int_{\mathcal{U}}\big[\delta_{A(-t)}\phi\cdot\mathcal{S}(\delta_{B(t)}\phi) + \delta_{B(-t)}\phi\cdot\mathcal{S}(\delta_{A(t)}\phi)\big] = \order{\hbar}.
    \label{Equation: consistency result}
\end{equation}

Now let us evaluate the commutator. From \eqref{Equation: expval AB 2}, we have
\begin{align}
    \expval*{\comm*{\hat A}{\hat B}} &= \expval*{\hat A \hat B} - \expval*{\hat B \hat A} \\
                         &= \frac{i\hbar}{2}\lim_{t\to 0}\int_{\mathcal{U}} \big[ \delta_{B(-t)}\phi \cdot\mathcal{S}(\delta_{A(t)}\phi) - \delta_{A(-t)}\phi \cdot\mathcal{S}(\delta_{B(t)}\phi) \big] + \order{\hbar^2}\\
                         &= -i\hbar\lim_{t\to 0}\int_{\mathcal{U}}\delta_{A(-t)}\phi \cdot\mathcal{S}(\delta_{B(t)}\phi) + \order{\hbar^2},
\end{align}
where we used \eqref{Equation: consistency result} to reach the third line. By the defining relation for the Poisson bracket \eqref{Equation: Poisson/commutator}, we may therefore take the $\hbar \to 0$ limit to obtain
\begin{equation}
    \pb{A}{B} = -\lim_{t\to 0}\int_{\mathcal{U}}\delta_{A(-t)}\phi \cdot\mathcal{S}(\delta_{B(t)}\phi) = \lim_{t\to 0}\delta_{A(-t)}B(t) = -\lim_{t\to 0}\delta_{B(-t)}A(t).
    \label{Equation: pb AB}
\end{equation}
There is one more useful way in which we may write the Poisson bracket. Using our causality assumptions, we have
\begin{equation}
    \pb{A}{B} = \pb{A}{B} - \lim_{t\to 0} \delta_{B(t)} A(-t)
              = \lim_{t\to 0} \big( \delta_{A(-t)}B(t) - \delta_{B(-t)}A(t) \big).
\end{equation}
We see that this Poisson bracket is essentially the same as the Peierls bracket~\cite{Peierls:10.2307/99080,DeWitt:1985bc,DeWitt:2003pm} restricted to observables on $\Sigma$. We may further use \eqref{Equation: delta W} to obtain
\begin{equation}
    \pb{A}{B} = \lim_{t\to 0} \int_{\mathcal{U}}\big[\delta_{B(t)}\phi\cdot\mathcal{S}(\delta_{A(-t)}\phi) - \delta_{A(-t)}\phi\cdot\mathcal{S}(\delta_{B(t)}\phi)\big].
\end{equation}
The right-hand side we recognise from \eqref{Equation: omega S}, which we may therefore use to write
\begin{equation}
    \pb{A}{B} = \lim_{t\to 0}\int_{\partial\mathcal{U}}\omega[\delta_{A(-t)}\phi,\delta_{B(t)}\phi].
\end{equation}

\subsection{Phase space and symplectic structure}

Let us write $\pb{A}{B} = \Pi(A,B)$. $\Pi$ is an antisymmetric bivector on configuration space known as the Poisson structure. We may view $\Pi$ as a map $A\mapsto \Pi(A)$ from observables on $\Sigma$ to vector fields on configuration space, defined by $\Pi(A)(B) = \Pi(A,B)$. $\Pi(A)$ is known as the Hamiltonian vector field associated to the observable $A$, and may be thought of as the field variation resulting from the application of $A$ as an operator. From \eqref{Equation: pb AB} it is clear that
\begin{equation}
    \Pi(A) = \lim_{t\to 0}\delta_{A(-t)}\phi.
\end{equation}
So this Poisson structure is in agreement with the saddlepoint approximation.

A standard result in Poisson geometry says that the commutator of any two Hamiltonian vector fields gives a third. By Frobenius' theorem, the Hamiltonian vector fields therefore span the tangent spaces to the leaves\footnote{We take the convention that each leaf only has one connected component.} of a regular foliation of the configuration space. These leaves are known as symplectic leaves. Let $\mathscr{P}_\Sigma$ be the symplectic leaf containing $\phi_0$. 

Suppose one starts at $\phi_0$ and applies some operators at $\Sigma$. This corresponds to flowing along some combination of Hamiltonian vector fields. During this flow, the state must remain in $\mathscr{P}_\Sigma$, because the Hamiltonian vector fields are tangent to $\mathscr{P}_\Sigma$. Conversely, it is possible to reach any field configuration in $\mathscr{P}_\Sigma$ by flowing along the appropriate Hamiltonian vector fields, i.e.\ by applying the right operators at $\Sigma$.

So $\mathscr{P}_\Sigma$ is the space of field configurations which can be explored by the application of operators at $\Sigma$. It therefore makes sense to use $\mathscr{P}_\Sigma$ as the phase space for the degrees of freedom on $\Sigma$. 

It remains to obtain a symplectic structure on $\mathscr{P}_\Sigma$. This is the inverse of $\Pi$ restricted to $\mathscr{P}_\Sigma$, which another standard result in Poisson geometry says is unique. Note that
\begin{equation}
    \delta A = -\int_{\mathcal{U}} \delta\phi \cdot \mathcal{S}(\Pi(A)).
\end{equation}
Therefore, the map
\begin{equation}
    \Omega(\hat\delta\phi) = -\int_{\mathcal{U}} \delta\phi \cdot \mathcal{S}(\hat\delta\phi) 
\end{equation}
inverts $\Pi$. Restricting this to $\mathscr{P}_\Sigma$, we have
\begin{equation}
    \Omega(\hat\delta\phi) = -\lim_{t\to 0}\int_{\mathcal{U}} \delta\phi(-t) \cdot \mathcal{S}(\hat\delta\phi(t)).
\end{equation}
In this expression, $\delta\phi(-t),\delta\phi(t)$ are field variations originating from operator insertions on $\Sigma(-t)$ and $\Sigma(t)$ respectively. Using causality, this may be written
\begin{equation}
    \Omega(\hat\delta\phi) = \lim_{t\to 0}\int_{\mathcal{U}}\big[ \hat\delta\phi(-t) \cdot\mathcal{S}(\delta\phi(t)) - \delta\phi(t) \cdot \mathcal{S}(\hat\delta\phi(-t))\big] = \lim_{t\to 0} \int_{\partial\mathcal{U}} \omega [\delta\phi(t),\hat\delta\phi(-t)].
\end{equation}
$\Omega$ may be viewed as a 2-form on $\mathscr{P}_\Sigma$:
\begin{equation}
    \Omega[\delta_1\phi,\delta_2\phi] = \lim_{t\to 0} \int_{\partial\mathcal{U}} \omega[\delta_1\phi(t),\delta_2\phi(-t)].
    \label{Equation: symplectic contour}
\end{equation}
This is the symplectic structure.

\subsection{Comparison to previous approach}
Let us hide the $t \to 0$ limit, and write the symplectic structure as\footnote{It is not always safe to assume this expression is valid, because of possible singular behaviour at $t = 0$. Nevertheless, it is at least heuristically useful.}
\begin{equation}
    \Omega[\delta_1\phi,\delta_2\phi] = \int_{\partial\mathcal{U}} \omega[\delta_1\phi,\delta_2\phi].
    \label{Equation: symplectic contour abuse}
\end{equation}
Recall the claim of Section \ref{Section: Review} for the value of the symplectic structure:
\begin{equation}
    \Omega[\delta_1\phi,\delta_2\phi] = \int_\Sigma \omega[\delta_1\phi,\delta_2\phi].
    \label{Equation: previous symplectic}
\end{equation}
The two expressions in \eqref{Equation: symplectic contour abuse} and \eqref{Equation: previous symplectic} are clearly very similar. They are both integrals of $\omega$, but over different surfaces. The integral in \eqref{Equation: symplectic contour abuse} is done over a contour around $\Sigma$, while the integral in \eqref{Equation: previous symplectic} is done over $\Sigma$ itself.

The reader might wonder whether \eqref{Equation: symplectic contour abuse} is logically equivalent to \eqref{Equation: previous symplectic}, i.e.\ whether one of these equations implies the other. This is not the case -- the expression in \eqref{Equation: previous symplectic} is sensitive to the ambiguity \eqref{Equation: omega ambiguity}, while the expression in \eqref{Equation: symplectic contour abuse} is not (since $\partial\mathcal{U}$ has no boundary). Since they do not share this quality, they must be logically distinct. Additionally, the space on which $\delta_1\phi,\delta_2\phi$ exist on is different in each expression. In \eqref{Equation: previous symplectic}, they are tangent vectors to $\mathscr{P}$, the space of on-shell field configurations. On the other hand, in \eqref{Equation: symplectic contour abuse}, they are tangent vectors to $\mathscr{P}_\Sigma$, which consists of all field configurations which can be obtained by applying operators at $\Sigma$. Such field configurations only need obey the equations of motion away from $\Sigma$.

The fact that \eqref{Equation: symplectic contour abuse} is insensitive to the ambiguity means that we have done what we set out to do in this section. The symplectic structure, and the theory of the degrees of freedom in the subregion, are now well-defined.

Note that it is possible to obtain an expression from \eqref{Equation: symplectic contour abuse} that more closely resembles \eqref{Equation: previous symplectic}, and we will now briefly outline how this would work. Suppose $\delta\phi$ is a field variation caused by operator insertions at $\Sigma$. By causality, $\delta\phi$ can only have support in $J^+(\Sigma)$, the causal future of $\Sigma$.
\begin{figure}[H]
    \centering
    \begin{tikzpicture}[thick,scale=1.6]
        \begin{scope}
            \clip (2.5,0) circle (2.5 and 0.5);
            \fill[blue,opacity=0.15] (1,0) -- (0,1) -- (5,1) -- (4,0) -- cycle;
            \draw[blue,dotted] (1,0) -- (0,1); 
            \draw[blue,dotted] (4,0) -- (5,1);
        \end{scope}
        \draw [decorate,decoration={brace,amplitude=10pt}] (0.7,0.5) -- (4.3,0.5);
        \node[above] at (2.5,0.7) {$\Sigma^+$};
        \draw (0.7,0.5) -- (0.7,0.4);
        \draw (4.3,0.5) -- (4.3,0.4);
        \draw[red] (1,0) -- (4,0) node[midway,below] {$\Sigma$};
        \draw (2.5,0) circle (2.5 and 0.5);
        \node[below] at (3.5,-0.5) {$\mathcal{U}$};
        \begin{scope}
            \clip (1,0) -- (0,1) -- (5,1) -- (4,0) -- cycle;
            \draw[line width = 1.5pt, blue] (2.5,0) circle (2.5 and 0.5);
        \end{scope}
    \end{tikzpicture}
    \caption{The support of $\delta\phi$ on $\partial\mathcal{U}$ is contained in $\Sigma^+=J^+(\Sigma)\cap\partial\mathcal{U}$.}
    \label{Figure: causal dependence}
\end{figure}
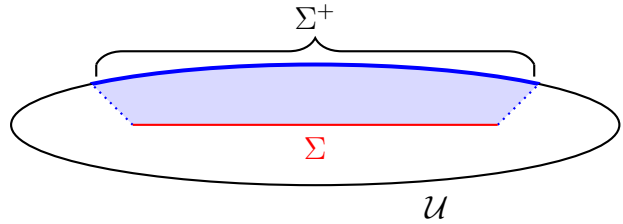
\noindent The $\delta_1\phi,\delta_2\phi$ appearing in \eqref{Equation: symplectic contour abuse} are two such field variations. Hence $\omega[\delta_1\phi,\delta_2\phi]$ also only has support in $J^+(\Sigma)$. Let $\Sigma^+=J^+(\Sigma)\cap\partial\mathcal{U}$. Na\"ively one might write
\begin{equation}
    \Omega[\delta_1\phi,\delta_2\phi] = \int_{\partial\mathcal{U}}\omega[\delta_1\phi,\delta_2\phi] = \int_{\Sigma^+} \omega[\delta_1\phi,\delta_2\phi].
\end{equation}
However, the right-hand side above cannot possibly be correct, because suddenly it is once more subject to the ambiguity \eqref{Equation: omega ambiguity}.

The reason this has happened is that we have failed to account for singular and distributional behaviour near $\partial\Sigma^+$, and splitting the integral up in this way only works when the integrand is sufficiently smooth. The proper way to carry out this split must involve some kind of regularisation at $\partial\Sigma$, where one integrates $\omega$ against appropriately chosen smooth test functions. The generic result of such a regularisation would be an expression of the form
\begin{equation}
    \Omega[\delta_1\phi,\delta_2\phi] = \int_{\Sigma^+}\omega[\delta_1\phi,\delta_2\phi] + \int_{\partial\Sigma^+} X[\delta_1\phi,\delta_2\phi],
    \label{Equation: regularised omega}
\end{equation}
where, under the ambiguity transformation \eqref{Equation: omega ambiguity}, $X$ transforms as
\begin{equation}
    X[\delta_1\phi,\delta_2\phi] \to X[\delta_1\phi,\delta_2\phi] + \delta_1(\alpha[\delta_2\phi]) - \delta_2(\alpha[\delta_1\phi]) - \alpha[\delta_{12}\phi].
\end{equation}
Such a transformation rule is necessary to ensure that $\Omega$ remains unaffected by the ambiguity.

\begin{figure}[H]
    \centering
    \begin{tikzpicture}[thick,scale=1.6]
        \begin{scope}
            \clip (2.5,0) circle (4.5 and 1);
            \fill[blue,opacity=0.15] (1,0) -- (-1,2) -- (6,2) -- (4,0) -- cycle;
            \draw[blue,dotted] (1,0) -- (0,1); 
            \draw[blue,dotted] (4,0) -- (5,1);
        \end{scope}
        \draw [decorate,decoration={brace,amplitude=10pt}] (0.2,1) -- (4.8,1);
        \node[above] at (2.5,1.2) {$\Sigma^+_n$};
        \draw (0.2,1) -- (0.2,0.9);
        \draw (4.8,1) -- (4.8,0.9);
        \draw[red] (1,0) -- (4,0) node[midway,below] {$\Sigma$};
        \draw (2.5,0) circle (2.5 and 0.36);
        \draw[black!70] (2.5,0) circle (3.5 and 0.68);
        \draw[black!40] (2.5,0) circle (4.5 and 1);
        \node[below] at (5.8,-0.7) {$\mathcal{U}_1$};
        \node[below] at (5.1,-0.45) {$\mathcal{U}_2$};
        \node[below] at (4.4,-0.2) {$\mathcal{U}_3$};
        \node[below] at (4.0,-0.05) {$\dots$};
        \begin{scope}[line width = 1.5pt, blue]
            \clip (1,0) -- (-0.2,1.2) -- (5.2,1.2) -- (4,0) -- cycle;
            \draw (2.5,0) circle (2.5 and 0.36);
            \draw[blue!70!white] (2.5,0) circle (3.5 and 0.68);
            \draw[blue!40!white] (2.5,0) circle (4.5 and 1);
        \end{scope}
    \end{tikzpicture}
    \caption{A sequence of $\mathcal{U}_n$ such that $\Sigma^+_n\to\Sigma$.}
    \label{Figure: sigma+ to sigma}
\end{figure}
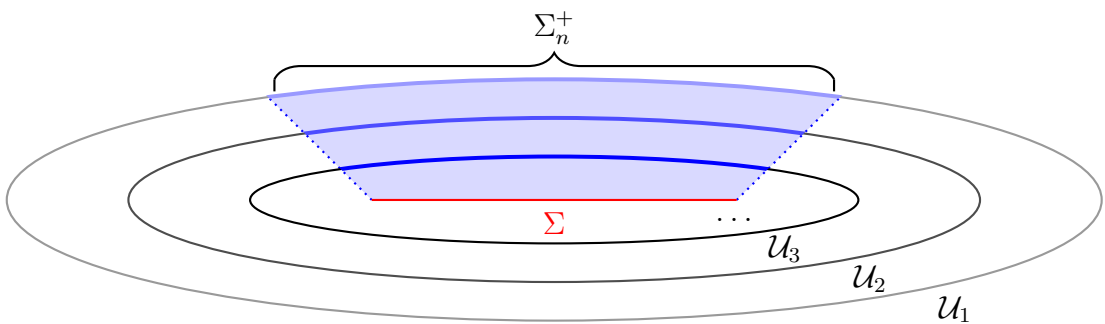

Once one has obtained \eqref{Equation: regularised omega}, one can proceed as follows. The expression in \eqref{Equation: symplectic contour abuse} is valid so long as $\mathcal{U}$ contains $\Sigma$. Let us consider a sequence of $\mathcal{U}\in \mathcal{U}_1,\mathcal{U}_2,\dots,\mathcal{U}_n,\dots$ that contain $\Sigma$, and let $\Sigma_n^+ = J^+(\Sigma)\cap\partial\mathcal{U}_n$. One can choose $\mathcal{U}_n$ such that $\Sigma_n^+\to\Sigma$ as $n\to \infty$. An example is given in Figure \ref{Figure: sigma+ to sigma}.
\noindent Once the $n\to \infty$ limit is taken, \eqref{Equation: regularised omega} takes the form
\begin{equation}
    \Omega[\delta_1\phi,\delta_2\phi] = \int_{\Sigma}\omega[\delta_1\phi,\delta_2\phi] + \int_{\partial\Sigma} X[\delta_1\phi,\delta_2\phi],
    \label{Equation: matched omega}
\end{equation}
where in this expression $\delta_1\phi$ and $\delta_2\phi$ should be evaluated `just to the future of' $\Sigma$.

The expression \eqref{Equation: matched omega} should be viewed as a version of \eqref{Equation: previous symplectic} for which the correct boundary term has been identified. It would be interesting to evaluate this boundary term directly and see whether it agrees with the boundary terms chosen in other contexts, for example in~\cite{Haco:2018ske} (although see Section \ref{Section: Gauge} for a reason that it can't possibly match that boundary term exactly). We leave exploration of this to future work. However, we will comment that it is not obvious that \eqref{Equation: matched omega} has any inherent advantages over \eqref{Equation: symplectic contour abuse}, other than its ease of comparison to previous studies. In fact, it is our view that the contour integral in \eqref{Equation: symplectic contour abuse} may be the more flexible expression. Whether this is actually true will hopefully become clear in the future.

Finally, we should note that even though we assumed $\Sigma$ was a partial Cauchy surface in the above, we can instead just assume that $\Sigma$ is a set containing no two points which can be connected by a causal curve. Then the steps in the above derivation all still follow, and \eqref{Equation: symplectic contour abuse} still applies (so long as $\mathcal{U}$ is chosen to contain $\Sigma$). For example, $\Sigma$ could be a codimension 2 submanifold, or it could not even be a manifold at all. In these cases there is still a notion of the degrees of freedom associated to $\Sigma$, and \eqref{Equation: symplectic contour abuse} provides a symplectic structure for these degrees of freedom. But \eqref{Equation: previous symplectic} clearly cannot be applied, as there is no well-defined way to integrate $\omega$ over such a set. Thus, \eqref{Equation: symplectic contour abuse} is applicable in a larger range of circumstances than \eqref{Equation: previous symplectic}.

\subsection{Asymptotic boundaries}

\changed{So far we have been discussing subregions with finite boundaries, in which case it is always possible to find a $\mathcal{U}$ which encloses the subregion. In the case that $\Sigma$ has an asymptotic boundary it is not clear that this can be done. However, our results do still apply in this case, subject to the following interpretation.}

\changed{It is important to recognise that an asymptotic boundary really only makes sense as the limit of a finite boundary as some parameter goes to infinity. For example, in an asymptotically flat spacetime, one can consider a Cauchy surface $\Sigma$ with an asymptotic boundary at spacelike infinity. But this is only really defined as the limit of $\Sigma_r$ as $r\to\infty$, where $\Sigma_r$ is a partial Cauchy surface whose boundary is a sphere of radius $r$.}

\changed{Such a prescription for an asymptotic boundary regularises many calculations which would otherwise not be well defined. For example, integrals over $\Sigma$ should really be considered as integrals over $\Sigma_r$, in the limit as $r\to\infty$:}
\begin{equation}
    \int_\Sigma = \lim_{r\to\infty} \int_{\Sigma_r}.
\end{equation}
\changed{Similarly, integrals over the asymptotic boundary should be viewed as integrals over the finite boundary $\partial\Sigma_r$, in the limit as $r\to\infty$:}
\begin{equation}
    \int_{\partial\Sigma} = \lim_{r\to\infty} \int_{\partial\Sigma_r}.
\end{equation}
\changed{Different choices of $\Sigma_r$ can lead to different integration results. This should not be viewed as an ambiguity in the definition of integration, but rather a dependence of the integral upon the definition of the asymptotic boundary over which it is being performed.}

\changed{Although it is not always made explicit, this affects \eqref{Equation: previous symplectic}. In particular, the symplectic structure given in Section \ref{Section: Review} has an integral in it, and so contains a limit as $r \to \infty$. To be explicit, we have}
\begin{equation}
    \int_\Sigma \omega = \lim_{r\to\infty} \int_{\Sigma_r}\omega.
\end{equation}

\changed{Armed with this realisation, it is clear how to extend our results to the case of an asymptotic boundary. For each value of $r$, we can find a $\mathcal{U}_r$ which encloses $\Sigma_r$. We can compute the symplectic structure associated with $\Sigma_r$ as an integral over $\partial\mathcal{U}_r$. Then we can take the limit as $r\to \infty$, obtaining}
\begin{equation}
    \Omega = \lim_{r\to\infty}\int_{\partial\mathcal{U}_r}\omega.
\end{equation}
\changed{Thus we get a well-defined symplectic structure for $\Sigma$, even though $\Sigma$ has an asymptotic boundary.}
\begin{figure}[H]
    \centering
    \begin{tikzpicture}[thick,scale=1.2]
        \draw[red] (0,0) -- (2,0) node[below, midway] {$\Sigma_r$};
        \draw (1,0) ellipse (2 and 1);
        \node[above] at (1,1) {$\mathcal{U}_r$};
        \draw[black!70,->] (-0.1,0) -- (-0.5,0);
        \draw[black!70,->] (2.1,0) -- (2.5,0);

        \draw[very thick,->] (3.75,0) -- (5.75,0) node[midway,above] {$r\to \infty$};

        \draw[red,dotted] (6.5,0) -- (7,0);
        \draw[red,dotted] (11.5,0) -- (12,0);
        \draw[dotted] (6.5,1) -- (7,1);
        \draw[dotted] (11.5,1) -- (12,1);
        \draw[dotted] (6.5,-1) -- (7,-1);
        \draw[dotted] (11.5,-1) -- (12,-1);
        \draw[red] (7,0) -- (11.5,0) node[below, midway] {$\Sigma$};
        \draw (7,1) -- (11.5,1) node[above, midway] {$\mathcal{U}$};
        \draw (7,-1) -- (11.5,-1);
    \end{tikzpicture}
    \caption{\changed{For $\Sigma$ with an asymptotic boundary, the symplectic structure is obtained by integrating over $\partial U_r$, and then taking the limit as $U_r$ grows to contain the entirety of $\Sigma$.}}
    \label{Figure: asymptotic}
\end{figure}
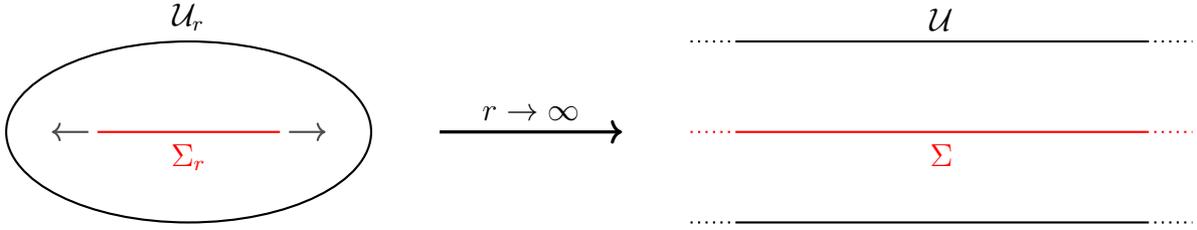

\section{Discussion}
\label{Section: Conclusions}

In this paper we have shown that the symplectic structure associated to a subregion with partial Cauchy surface $\Sigma$ should be written as the contour integral of the form $\omega$ around $\Sigma$. This is in contrast to the previous notion that the symplectic structure should be written as the integral of $\omega$ over $\Sigma$ itself, and, as we have discussed, resolves the boundary ambiguities inherent to that belief. There are a number of other immediate consequences of our results, a couple of which we will briefly describe below. A more full exploration of these topics will be the subject of forthcoming work.

\subsection{Gauge symmetries}
\label{Section: Gauge}

Suppose that the field theory we are interested in has a gauge symmetry, and consider a gauge transformation $\phi\to\phi+\delta_\lambda\phi$, where $\lambda$ is some parameter. One may show that, when the equations of motion are obeyed, $\omega[\delta\phi,\delta_\lambda\phi]$ is an exact form~\cite{Wald:2010:doi:10.1063/1.528839}. We will write $\omega[\delta\phi,\delta_\lambda\phi] = \dd(\slashed\delta q_\lambda[\delta\phi])$. If we assume that the symplectic structure is given by $\Omega = \int_\Sigma\omega$, and that the phase space consists of solutions to the equations of motion, then we clearly have
\begin{equation}
    \Omega[\delta\phi,\delta_\lambda\phi] = \int_\Sigma \omega[\delta\phi,\delta_\lambda\phi] = \int_{\partial\Sigma} \slashed\delta q_\lambda,
\end{equation}
and in general, the integral on the right-hand side will not vanish. The implication is that there are some gauge transformations which do not correspond to degenerate directions of the symplectic structure. Such gauge transformations have non-trivial action at $\partial\Sigma$, and are referred to as large. One is forced to conclude that large gauge transformations have observable consequences, and therefore that there must exist physically relevant gauge-dependent degrees of freedom. This is a slightly confusing conclusion, because one might have expected that the theory was constructed in such a way that this is not the case. Nevertheless, it has been taken seriously in many studies, and numerous subtle arguments have been made for its validity, for example~\cite{Blommaert:2018oue}.

However, as we have shown, the correct expression for the symplectic structure is $\Omega=\int_{\partial\mathcal{U}}\omega$. The correct phase space only includes configurations which disobey the equations of motion at $\Sigma$, and $\Sigma\cap\partial\mathcal{U}=\emptyset$, so the equations of motion are obeyed at $\partial\mathcal{U}$. Therefore, we have
\begin{equation}
    \Omega[\delta\phi,\delta_\lambda\phi] = \int_{\partial\mathcal{U}} \dd(\slashed\delta q_\lambda) = 0,
\end{equation}
where the last equality holds because $\partial\mathcal{U}$ has no boundary. Therefore, we can unequivocally state that \emph{all} gauge transformations are non-physical, even large ones. This much more closely fits our intuition for what a gauge transformation is.

We want to emphasise that we do not claim that this result invalidates previous work on large gauge transformations. Rather, we take the view that it should change the interpretation of that work. To be clear what we mean, suppose that there is a $Q_\lambda$ such that
\begin{equation}
    \delta Q_\lambda = \int_{\partial\Sigma} \slashed\delta q_\lambda.
\end{equation}
If one assumes that $\Omega=\int_\Sigma\omega$, then the gauge transformation corresponding to $\lambda$ is integrable, and generated by $Q_\lambda$. Suppose however that instead we use the correct symplectic structure $\Omega = \int_{\partial\mathcal{U}}\omega$. Then $Q_\lambda$ can still be thought of as the generator of some field transformation $\phi \to \phi + \tilde\delta_\lambda\phi$ obeying
\begin{equation}
    \Omega[\delta\phi,\tilde\delta_\lambda\phi] = \int_{\partial\mathcal{U}} \omega[\delta\phi,\tilde\delta_\lambda\phi] = \delta Q_\lambda.
\end{equation}
Clearly, $\tilde\delta_\lambda\phi$ cannot be merely a gauge transformation, since we have just shown that all gauge transformations are degenerate in the symplectic structure. We will refer to $\tilde\delta_\lambda\phi$ as a pseudo-gauge transformation, due to its subtle similarity with a true gauge transformation\footnote{It would be interesting to compare these pseudo-gauge transformations to the `would-be' gauge transformations of~\cite{Blommaert:2018oue}.}. We believe that the pseudo-gauge transformations are worth studying, and suspect that many of the results regarding large gauge transformations should instead be interpreted as applying to pseudo-gauge transformations. 

\subsection{Correlations between distinct subregions}

Consider the subregions associated to two distinct and spatially separated partial Cauchy surfaces $\Sigma_1,\Sigma_2$. Let us ask the following question: can an observation in one of these subregions affect an observation in the other? Let $A_1,A_2$ be observables on $\Sigma_1,\Sigma_2$, generating field variations $\delta_1\phi,\delta_2\phi$ respectively. We can answer our question by calculating the Poisson bracket of $A_1$ and $A_2$, which is equal to $\Omega[\delta_1\phi,\delta_2\phi]$. 

\begin{figure}[H]
    \centering
    \begin{tikzpicture}[thick,scale=1.6]
        \begin{scope}
            \clip (4,0.2) circle (4 and 0.8);
            \fill[blue,opacity=0.15] (1,0) -- (0,1) -- (4.5,1) -- (3.5,0) -- cycle;
            \fill[blue,opacity=0.15] (4.5,0) -- (3.5,1) -- (8,1) -- (7,0) -- cycle;
            \draw[blue,dotted] (1,0) -- (0,1); 
            \draw[blue,dotted] (3.5,0) -- (4.5,1);
            \draw[blue,dotted] (4.5,0) -- (3.5,1);
            \draw[blue,dotted] (7,0) -- (8,1);
        \end{scope}
        \draw[red] (1,0) -- (3.5,0) node[midway,below] {$\Sigma_1$};
        \draw[red] (4.5,0) -- (7,0) node[midway,below] {$\Sigma_2$};
        \draw (4,0.2) circle (4 and 0.8);
        \node[below] at (4,-0.6) {$\mathcal{U}$};
        \begin{scope}
            \clip (4.5,0) -- (3.3,1.2) -- (8,1) -- (7,0) -- cycle;
            \clip (1,0) -- (0,1) -- (4.7,1.2) -- (3.5,0) -- cycle;
            \draw[line width = 1.5pt, blue] (4,0.2) circle (4 and 0.8);
        \end{scope}
        \node[above,blue] at (2.25,0.2) {$\delta_1\phi$};
        \node[above,blue] at (5.75,0.2) {$\delta_2\phi$};
        \draw [decorate,decoration={brace,amplitude=10pt}] (3.5,1.05) -- (4.5,1.05);
        \node[above] at (4,1.25) {$\partial\mathcal{U}\cap J^+(\Sigma_1)\cap J^+(\Sigma_2)$};
    \end{tikzpicture}
    \caption{The joint support of $\delta_1\phi,\delta_2\phi$ on $\partial\mathcal{U}$ is contained in $\partial\mathcal{U}\cap J^+(\Sigma_1)\cap J^+(\Sigma_2)$.}
    \label{Figure: two subregions gap}
\end{figure}
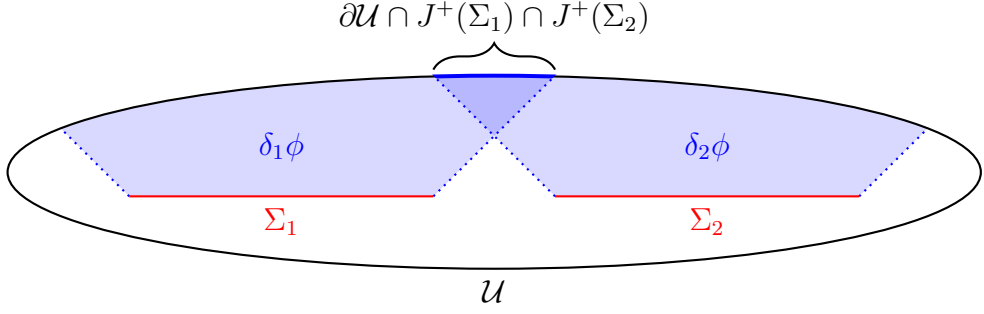

There are two cases to consider. In the first case, we assume there is a `gap' between $\Sigma_1$ and $\Sigma_2$. Let us pick a $\mathcal{U}$ as in Figure \ref{Figure: two subregions gap}.
\noindent By causality, the supports of $\delta_1\phi,\delta_2\phi$ must be contained in $J^+(\Sigma_1),J^+(\Sigma_2)$ respectively. Therefore, the integrand of $\Omega[\delta_1\phi,\delta_2\phi] = \int_{\partial\mathcal{U}}\omega[\delta_1\phi,\delta_2\phi]$ can only be supported in $\partial\mathcal{U}\cap J^+(\Sigma_1)\cap J^+(\Sigma_2)$.

\begin{figure}[H]
    \centering
    \begin{tikzpicture}[thick,scale=1.6]
        \begin{scope}
            \clip (4,-0.1) circle (4 and 0.5);
            \fill[blue,opacity=0.15] (1,0) -- (0,1) -- (4.5,1) -- (3.5,0) -- cycle;
            \fill[blue,opacity=0.15] (4.5,0) -- (3.5,1) -- (8,1) -- (7,0) -- cycle;
            \draw[blue,dotted] (1,0) -- (0,1); 
            \draw[blue,dotted] (3.5,0) -- (4.5,1);
            \draw[blue,dotted] (4.5,0) -- (3.5,1);
            \draw[blue,dotted] (7,0) -- (8,1);
        \end{scope}
        \draw[red] (1,0) -- (3.5,0) node[midway,below] {$\Sigma_1$};
        \draw[red] (4.5,0) -- (7,0) node[midway,below] {$\Sigma_2$};
        \draw (4,-0.1) circle (4 and 0.5);
        \node[below] at (4,-0.65) {$\mathcal{U}$};
    \end{tikzpicture}
    \caption{With this choice of $\mathcal{U}$, we have $\partial\mathcal{U}\cap J^+(\Sigma_1)\cap J^+(\Sigma_2)=\emptyset$.}
    \label{Figure: two subregions gap 2}
\end{figure}
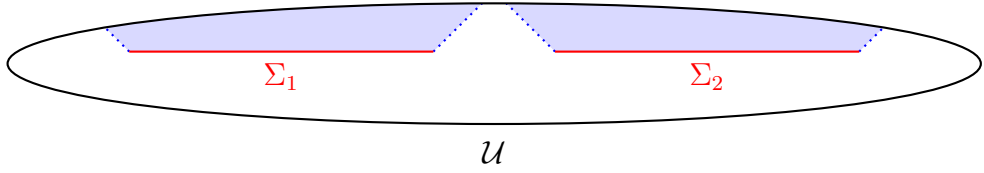

We could at this point try to compute this integral directly, but it turns out to be much easier to instead just pick a different $\mathcal{U}$ such that $\partial\mathcal{U}\cap J^+(\Sigma_1)\cap J^+(\Sigma_2) = \emptyset$, as in Figure \ref{Figure: two subregions gap 2}.
\noindent This choice is made possible by the gap between $\Sigma_1$ and $\Sigma_2$. The support of the integrand in $\Omega[\delta_1\phi,\delta_2\phi] = \int_{\partial\mathcal{U}}\omega[\delta_1\phi,\delta_2\phi]$ is empty, so $\Omega[\delta_1\phi,\delta_2\phi] = 0$. Therefore, the Poisson bracket of $A_1$ and $A_2$ vanishes, and we may conclude that no observation on $\Sigma_1$ can affect an observation on $\Sigma_2$.

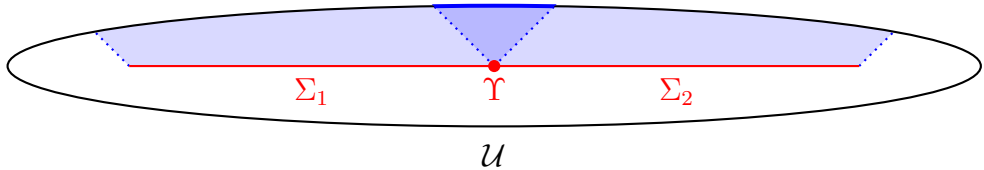
\begin{figure}[H]
    \centering
    \begin{tikzpicture}[thick,scale=1.6]
        \begin{scope}
            \clip (4,0) circle (4 and 0.5);
            \fill[blue,opacity=0.15] (1,0) -- (0,1) -- (5,1) -- (4,0) -- cycle;
            \fill[blue,opacity=0.15] (4,0) -- (3,1) -- (8,1) -- (7,0) -- cycle;
            \draw[blue,dotted] (1,0) -- (0,1); 
            \draw[blue,dotted] (4,0) -- (5,1);
            \draw[blue,dotted] (4,0) -- (3,1);
            \draw[blue,dotted] (7,0) -- (8,1);
        \end{scope}
        \draw[red] (1,0) -- (4,0) node[midway,below] {$\Sigma_1$};
        \draw[red] (4,0) -- (7,0) node[midway,below] {$\Sigma_2$};
        \fill[red] (4,0) circle (0.05) node[below] {$\Upsilon$};
        \draw (4,0) circle (4 and 0.5);
        \node[below] at (4,-0.55) {$\mathcal{U}$};
        \begin{scope}
            \clip (4,0) -- (3,1) -- (8,1) -- (7,0) -- cycle;
            \clip (1,0) -- (0,1) -- (5,1) -- (4,0) -- cycle;
            \draw[line width = 1.5pt, blue] (4,0) circle (4 and 0.5);
        \end{scope}
    \end{tikzpicture}
    \caption{When there is no gap, it is not possible to choose $\mathcal{U}$ such that $\partial\mathcal{U}\cap J^+(\Sigma_1)\cap J^+(\Sigma_2)=\emptyset$. The red dot denotes the edge $\Upsilon$ joining $\Sigma_1$ and $\Sigma_2$, commonly known as the entangling surface.}
    \label{Figure: two subregions no gap}
\end{figure}

The situation changes when there is no gap between $\Sigma_1$ and $\Sigma_2$, which is the second case we consider, and is shown in Figure \ref{Figure: two subregions no gap}. It is no longer possible in such circumstances to choose $\mathcal{U}$ such that $\partial\mathcal{U}\cap J^+(\Sigma_1)\cap J^+(\Sigma_2)=\emptyset$, so we can not use the above trick to show that $\Omega[\delta_1\phi,\delta_2\phi]=0$.
\noindent In fact, it sometimes is possible to find $A_1$ and $A_2$ such that $\pb{A_1}{A_2} = \Omega[\delta_1\phi,\delta_2\phi] \ne 0$. Hence there \emph{are} observations on $\Sigma_1$ which can affect observations on $\Sigma_2$, even though these two surfaces are distinct and spatially separated. 

One usually postulates that operators supported in spatially separate subregions must commute. Our derivation of the Poisson bracket seems to require that one interpret $\pb{A_1}{A_2}$ as the classical limit of a commutator of such operators -- but the above observation suggests that this does not vanish. Thus there is a tension between the operator interpretation of the classical observables, and the postulate of commuting spatially separated operators. Since the postulate is, in our opinion, fundamental, we must conclude that the operator interpretation is not completely correct, and that the subregion observables are the classical limit of some more subtle manipulations of the quantum state.

Despite this issue, the Poisson bracket $\pb{A_1}{A_2}$ is still a function on phase space that measures the degree to which observations in one subregion affect observations in the other. Let $\mathscr{E}$ be the space of all such functions, i.e.\ the Poisson brackets of all possible observables $A_1$ on $\Sigma_1$ and $A_2$ on $\Sigma_2$. For a given field configuration, the functions in $\mathscr{E}$ take certain values, which can be thought of as parametrising the correlations between $\Sigma_1$ and $\Sigma_2$. These values only depend on the field configuration in the `interface' region $J^+(\Sigma_1)\cap J^+(\Sigma_2)$. This region is equivalent to $J^+(\Upsilon)$, where $\Upsilon = \overline{\Sigma}_1\cap\overline{\Sigma}_2$ is the edge joining $\Sigma_1$ with $\Sigma_2$ (here $\overline{\Sigma}_1$, $\overline{\Sigma}_2$ denote the closures of $\Sigma_1$, $\Sigma_2$ respectively). Thus, this correlation parametrisation should be thought of as depending on `edge modes' living on $\Upsilon$.

In quantum field theory, correlations between spatially separate subregions can only arise from entanglement between the quantum states in these subregions. Thus, we are led to believe that the ideas described here could enable a quantitative description of the entanglement configuration between the two subregions in terms of emergent degrees of freedom living at $\Upsilon$ (which is incidentally sometimes known as the entangling surface). This is a topic that has been the subject of much recent interest -- see~\cite{Harlow:2014yka,VanRaamsdonk:2016exw,Harlow:2018fse,Calabrese:2004eu,Ryu:2006bv,Ryu:2006ef,Hubeny:2007xt,VanRaamsdonk:2010pw,Lewkowycz:2013nqa,Donnelly:2016auv,Speranza:2017gxd,Kirklin:2018gcl,Camps:2018wjf,Dong:2018seb}, amongst many others. The details of this description obviously need to be worked out, but we are hopeful that this line of thought proves fruitful.

\section*{Acknowledgements}
\addcontentsline{toc}{section}{\protect\numberline{}Acknowledgements}
I am grateful to Malcolm Perry for many useful discussions. This work was supported by a grant from STFC.

\printbibliography

\end{document}